# High-speed varifocal optical-resolution photoacoustic microscope to achieve large depth of field based on a tunable acoustic gradient lens

XIANLIN SONG

School of Information Engineering, Nanchang University, Nanchang, China

**We developed an optical-resolution photoacoustic microscope (OR-PAM) capable of changing focal plane with high-speed by using a tunable acoustic gradient (TAG) lens. In our system, a TAG lens is designed to continuously changing the focal plane of OR-PAM by modulating its refractive power with fast-changing ultrasonic standing wave. The raster-scanning of the microscope and laser pulses are synchronized to the same phase of the driving signal of TAG lens. By selecting the synchronized phase, arbitrary focal plane can be chosen in a range of about 1 mm in 1.4 μs in our system. A phantom composed by tungsten wires is used to verified the varifocal capability of the system. *In-vivo* study was carried out, to further demonstrate the large depth-of-field (DoF) achieved by our method without any moving part.**

Photoacoustic imaging is a promising technique that combines optical contrast with ultrasonic detection to map the distribution of the absorbing pigments in biological tissues [1-5]. It has been widely used in biological researches, such as structural imaging of vasculature [6], brain structural and functional imaging [7], and tumor detection [8]. Considering the lateral resolution of photoacoustic microscopy (PAM), it can be classified into two categories: optical-resolution (OR-) and acoustic-resolution (AR-) PAM [9, 10]. In AR-PAM, the spatial resolution is determined by the acoustic focus, since the laser light is weekly or even not focused on the sample. Conversely, in the OR-PAM, the laser light is tightly focused into the sample to achieve sharp excitation. However, in OR-PAM, the DoF is quite limited, for it is determined by the optical condenser and is closely related to the optical focusing. The small DoF will prevent OR-PAM to achieve high-quality 3D images or acquire dynamic information in depth direction.

Depth scanning using motorized stage is widely used to address this issue as it is the most convenient method [11, 12]. To avoid slow mechanical scanning in the depth direction, several methods by engineering the illumination have been proposed. Double illumination can double the DoF by illuminating the sample from both top and bottom sides simultaneously. But it is only valid in transmission-mode OR-PAM [13]. Utilizing chromatic aberration of non-achromatic objective, multi-wavelength laser can generate multi-focus along the depth direction [14]. However, this method sacrifices the capability of functional imaging. Non-diffraction beam inherently own a large DoF. Bessel beam based PAM can extend the DoF to 1 mm with a lateral resolution of 7 μm [15]. In this system, a non-linear method must be used to suppress the artifacts introduced by the side lobes of the Bessel beam, which made it hard-to-use in the *in-vivo* imaging. Electrically tunable lens (ETL) has also been introduced in OR-PAM [16], the focal plane can be settled in about 15 ms (EL-10-30，Optotune AG). It is fast enough for pulsed lasers with a repetition rate of hundred-hertz, while being quite slow for those lasers with repetition rate of hundreds of kilo-hertz [11].

In this letter, we report a novel OR-PAM that can achieve large DoF by selection of focal planes using a high-speed TAG lens. The varifocal performance of our system was shown with a phantom made by tungsten wires. A zebrafish was also imaged to demonstrate the feasibility of our system in biomedical imaging.

Fig. 1 shows the scheme of the system. The system equipped with an Nd: YLF laser (IS8II-E, EdgeWave GmbH) irradiate laser light at the wavelength of 523 nm and a pulse repetition rate of 1 KHz. The laser beam is reshaped by an iris with a diameter of 0.8 mm and then focused by a plano-convex lens. A 50 μm-diameter pinhole placed after the focus is use as a spatial filter. Then the defocused laser beam is transformed into a collimated beam by a plano-convex lens. An objective lens is used to form a focus on the tip of a single-mode fiber. The light comes out of the distal end of the fiber is inserted into a fiber port (PAFA-X-4-A, Thorlabs)which transforms the guided laser beam into collimated beam with an output $1/e^2$ waist diameter of 0.65 mm. Then the optical axis of the laser beam is turned to vertical to reduce the gravity induced Y-coma aberration of the TAG lens.

The light come from TAG is delivered into a beam expander formed by plano-convex lenses L4 (f = 18 mm) and L5 (f = 150 mm). Hence the rear plane of TAG lens is conjugated to the back focal plane of the objective (4× Olympus objective, N.A. 0.14) with a magnification factor of 8.3. A home-made acoustic lens (NA=0.5) is glued on a ultrasound transducer (central frequency 50MHz, V214-BB-RM, Olympus) to detect the ultrasound [17]. The sample is placed on a glass slide. To couple the photoacoustic signals, we use a water tank. Both of sample and water tank are mounted on a 3D scanning stage which is assembled by a 2D linear stage (ANT95-XY, Aerotech) and a lifting stage (M-Z01.5G0, PhysikInstrumente). The 2D linear stage is used for the raster scanning of the sample. Photoacoustic signals detected by the ultrasound transducer is amplified by a amplifier (AU-1291, MITEQ) and acquired by a data acquisition card (ATS9350, Alazartech). In order to keep the scanning at the expected focal plane, the laser pulse, scanner, and TAG lens are synchronized. A sinusoidal signal with a frequency of 707 kHz generated by a function generator (DS345, Stanford Research Systems) drives the TAG lens at an eigenmode. The Sync Output of the function generator, which provides the synchronizing TTL square wave, is connected to the clock input of a D-type flip-flop (SN74AUC1G74, Texas Instruments). Position synchronized output (PSO) signal of the 2D scanner is sent to the data input. A Digital Delay and Pulse Generator (DG535, Stanford Research Systems) delays the output signal from the flip-flop with desired time for laser pulse triggering. Therefore, the PAM can work with arbitrary designated focal plane. And we can achieve focus-shifting by changing the delay time.

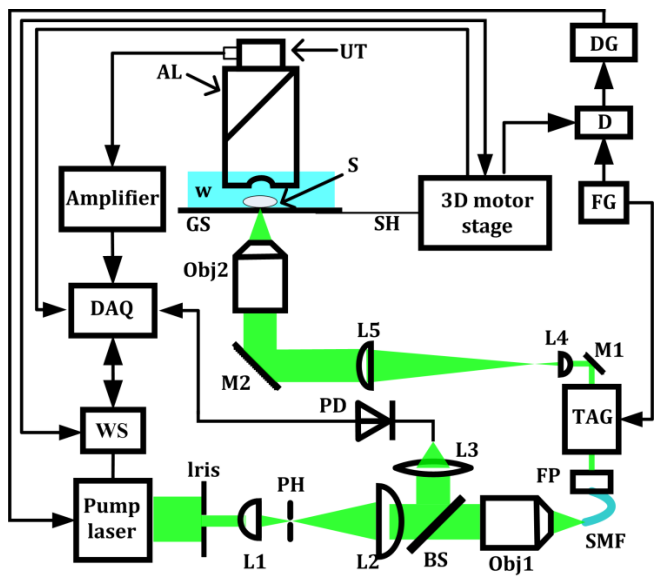

Fig. 1. Scheme of the varifocal OR-PAM system. AL, acoustic lens; BS, beam sampler; DAQ, data acquisition card; DG, Digital Delay and Pulse Generator ; D, D type flip-flop;FP, fiber port; FG, function generator; GS, glass slide; M1 and M2, mirrors; L1, L2, L3, L4 and L5, optical lenses; Obj1 and Obj2, objectives; PD, photodiode; PH, pinhole; S, sample; SH, sample holder; SMF, single mode fiber; TAG, TAG lens; UT, ultrasound transducer; W, water tank; WS, work station.

The home-made TAG lens used in our system consists of a cylindrical piezoelectric shell (PZT-8, Boston Piezo Optics) filled with a transparent silicone oil (100 cS, Sigma-Aldrich), with a refractive index of 1.403 and a speed of sound of 1000 m s$^{-1}$. As shown in Fig. 2(a), we validated the TAG lens with the similar synchronization paradigm used in our system, except that the PSO signal of the scanner is replaced with 1 kHz square wave from FG2 (TFG2006V,SUING). During the measurement, a 707 kHz sinusoidal signal generated by FG1 (DS345, Stanford Research Systems) drives the TAG lens. The wavefront is measured with a wavefront sensor (WFS150-5C, Thorlabs) placed 50 mm after the TAG lens output window. According to the correlation between lens power and radius of curvature (ROC) given in [18], the ROC can be calculated. By changing the delay time of the Digital Delay and Pulse Generator with the step of 100 ns , time-depended lens power is profiled. Fig. 2 (b) shows focusing power of TAG lens as a function of time with a 10 V$_{p-p}$ driving signal at a frequency of 707 kHz.

We then characterized lateral resolution of our system. During the test of lateral resolution , there was no signal driven on the TAG lens. The sharp edge of a cover glass (300 μm thick) which was evenly coated with red ink, was used to quantify the lateral resolution. The ESF (edge spread function) was extracted from a B scan image. Full-width-at-half-maximum (FWHM) of the LSF (line spread function) calculated as the derivative of the ESF，was 3.40 μm, as shown in Fig. 2 (c).

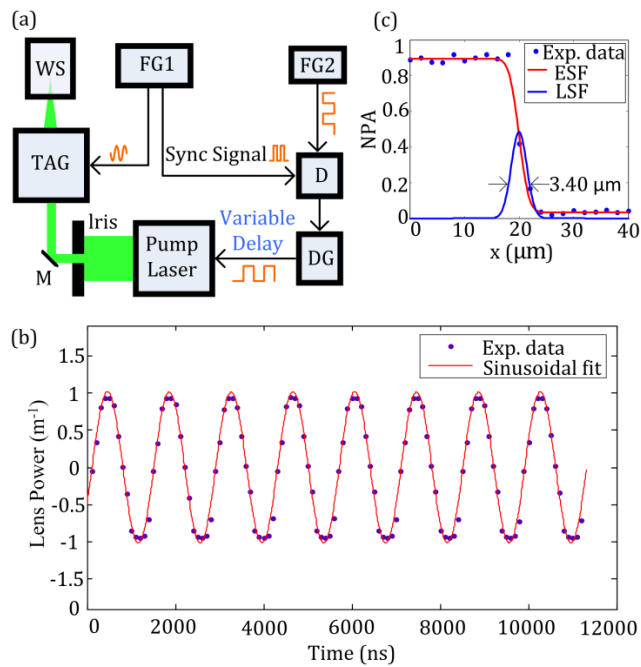

Fig. 2. (a) Schematic diagram for determining lens power of TAG lens using a wavefront sensor. (b) TAG lens focusing power as a function of time for a 10 Vp-p driving signal at a frequency of 707 kHz. The solid curve shows a two-parameter (amplitude and phase) sinusoidal fit to the data. (c) Lateral resolution of the system, edge spread function (ESF) extracted from the sharp edge of a glass slide and line spread function (LSF) obtained by taking the derivative of the ESF. DG, Digital Delay and Pulse Generator; D, D type flip-flop; FG1 and FG2, function generator; M, mirrors; TAG, tunable acoustic gradient index of refraction lens; WS, wavefront sensor. NPA, normalized photoacoustic amplitude.

We characterized the varifocal capability of our system. A network consists of three tungsten wires with diameters of 20 μm was used as the imaging phantom. The three tungsten wires were placed at three different layers in depth with a distance of about 500 μm between adjacent two of them. At first, no driven signal was loaded on the TAG lens. Then we fix the focal plane of our system at the middle layer, a maximum amplitude projection (MAP) PA image was acquired by raster scan, one tungsten wire was clearly resolved as shown in Fig. 3. (b). Next, TAG lens was driven with a 10 V$_{p-p}$ sinusoidal signal with a frequency of 707 kHz. By setting the delay time to 987 and 250 ns, MAP images of the other wires were also obtained and given in Fig. 3 (a) and (c), respectively. Fig. 3 (e), (f), and (g) shows the cross-sectional B scan image at the position indicated by the white lines in Fig. 3 (a), (b), and (c), respectively. It can be seen that while the in-plane tungsten wire can be clearly observed, the other wires are hardly identified. Finally, a data set was obtained by summing the three sets acquired above point-to-point. A synthetic B scan image is shown in Fig. 3(h), and the corresponding MAP image is shown in Fig. 3(d), with the whole tungsten wire network being recognized. Thus, large DoF PA imaging is achieved without any mechanical scanning. The varifocal range from the top of the tungsten wire network to the bottom was about 1 mm.

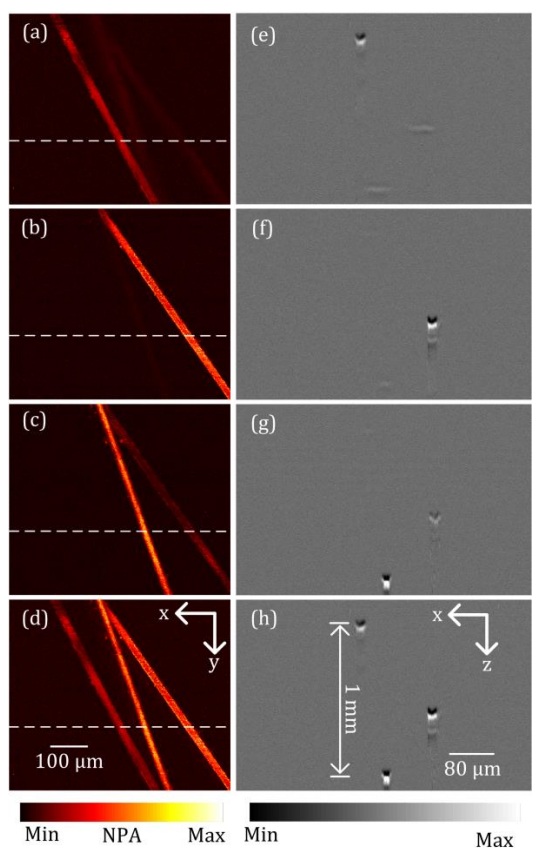

Fig. 3. Varifocal capability of the built system. (a), (b), and (c) are the MAP images of the tungsten wire network when the focal plane is located at different places, respectively. (d) is the MAP image by summing the three datasets for (a), (b) and (c). (e), (f), (g), and (h) is the B scan image whose position is indicated by the white dash line in (a), (b), (c), and (d), respectively. NPA, normalized photoacoustic amplitude.

We also showed that this system can be used for imaging of biological samples. A 30-days-old zebrafish (AB strain) was chosen for *in-vivo* imaging. Before the imaging, low melting point agarose (A-4018, Sigma-Aldrich) was dissolved at 40 °C (1.2% w/v), a culture dish was previously coated with a thin layer of the melted liquid agarose. When the temperature of liquid agarose dropped to 37 °C, a zebrafish was placed in the culture dish, lightly covered with the melted liquid agarose and oriented so that it was lying on its back waiting for polymerization. After the polymerization, some double-deionized water was pour into the culture dish. The temperature was kept around 25 °C during the imaging. Except for the delay times, the other parameters are the same with the phantom experiment.

TAG lens was firstly turned off and the pigments on the head of the zebrafish was clearly visualized in Fig. 4 (a), while the eyes part is quiet blurred. Then the TAG lens was turned on and the delay time was set to 1050 ns. The eyes were imaged with high resolution at this time, with no pigments being distinguished on the head of the zebrafish, as shown in Fig. 4(b). After we sum the two data sets acquired at both focal planes, all of the details can be found in Fig. 4(c).

It is worth noting that in our system we did not fully explore the high-speed tuning property of TAG lens, since the repetition rate of our laser is limited to 5 kHz, much lower than the driven signal of TAG lens. It is possible to achieve axial scan in one cycle of driven signal of TAG lens when we use a high-repetition rate laser [11] or a low-frequency driven signal of TAG lens [19]. Furthermore, several lasers with low repetition rate can also work together to fire several laser pulse in one cycle of driven signal of TAG lens.

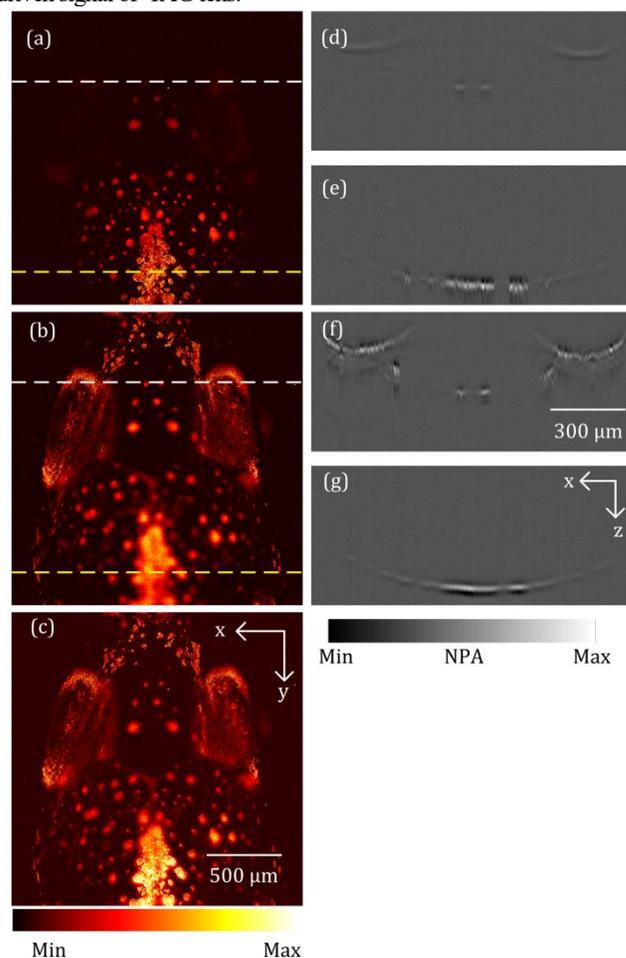

Fig. 4. Images of the head of a zebrafish acquired with our system. (a) and (b) are the MAP images of the head of a zebrafish when the pigments and eyes were in-plane, respectively. (c) is the MAP image of the summed data set provided by (a) and (b). (d) and (f) is the B scan image whose position is indicated by the white dash line in (a) and (b), respectively. (e) and (g) is the B scan image whose position is indicated by the yellow dash line in (a) and (b), respectively. NPA, normalized photoacoustic amplitude.

In summary, We developed a varifocal photoacoustic microscope with tunable focal plane based on a TAG lens. The varifocal range can reach 1 mm in axial when a 10 $V_{p-p}$ sinusoidal signal with frequency of 707 kHz was driven on the TAG lens. The focal plane can be changed continuously in the varifocal range by adjusting the delay time (delay phase) relative to the driven signal of TAG lens. The varifocal capability of our system could be used for axial scanning which may fulfill different requirements in biomedical researches.

**References**

1. L. V. Wang, and S. Hu, "Photoacoustic Tomography: In Vivo Imaging from Organelles to Organs," Science 335, 1458-1462 (2012).
2. P. Beard, "Biomedical photoacoustic imaging," Interface Focus 1, 602-631 (2011).